\documentclass[conference]{IEEEtran}
\IEEEoverridecommandlockouts
\usepackage{cite}
\usepackage{amsmath,amssymb,amsfonts}
\usepackage{algorithm}
\usepackage{algorithmic}
\usepackage{graphicx}
\usepackage{textcomp}
\usepackage{xcolor}
\def\BibTeX{{\rm B\kern-.05em{\sc i\kern-.025em b}\kern-.08em
    T\kern-.1667em\lower.7ex\hbox{E}\kern-.125emX}}
\begin{document}

\title{Optimizing Resource Allocation and Energy Efficiency in Federated Fog Computing for IoT}

\author{
\IEEEauthorblockN{Taimoor Ahmad\IEEEauthorrefmark{1}\IEEEauthorrefmark{3}, Anas Ali\IEEEauthorrefmark{2}}
\IEEEauthorblockA{\IEEEauthorrefmark{2}Department of Computer Science, National University of Modern Languages, Lahore, Pakistan}
\IEEEauthorblockA{\IEEEauthorrefmark{3}Department of Computer Science, The Superior University, Lahore, Pakistan\\
 anas.ali@numl.edu.pk, taimoor.ahmad1@superior.edu.pk}
}

\maketitle

\begin{abstract}
Address Resolution Protocol (ARP) spoofing attacks severely threaten Internet of Things (IoT) networks by allowing attackers to intercept, modify, or block communications. Traditional detection methods are insufficient due to high false positives and poor adaptability. This research proposes a multi-layered machine learning-based framework for intelligently detecting ARP spoofing in IoT networks. Our approach utilizes an ensemble of classifiers organized into multiple layers, each layer optimizing detection accuracy and reducing false alarms. Experimental evaluations demonstrate significant improvements in detection accuracy (up to 97.5\%), reduced false positive rates (less than 2\%), and faster detection time compared to existing methods. Our key contributions include introducing multi-layer ensemble classifiers specifically tuned for IoT networks, systematically addressing dataset imbalance problems, introducing a dynamic feedback mechanism for classifier retraining, and validating practical applicability through extensive simulations. This research enhances security management in IoT deployments, providing robust defenses against ARP spoofing attacks and improving reliability and trust in IoT environments.
\end{abstract}


\section{Introduction}

The Internet of Things (IoT) has become a transformative technology, reshaping how devices interact, communicate, and share data across numerous applications, including healthcare, industrial automation, smart homes, and smart cities. IoT networks integrate billions of interconnected devices, each generating and consuming vast amounts of data. While IoT brings unprecedented convenience, efficiency, and innovation, it simultaneously introduces substantial security vulnerabilities. Among these vulnerabilities, Address Resolution Protocol (ARP) spoofing represents a significant threat due to its ease of execution and potentially catastrophic impacts on IoT network security \cite{kumar2021survey}.

ARP spoofing is a form of cyber-attack where an attacker sends forged ARP messages onto a local area network. The objective is to link the attacker's MAC address with the IP address of another host, causing traffic intended for the legitimate host to be incorrectly routed to the attacker \cite{ouaddah2021towards}. Consequently, this allows attackers to intercept, modify, or block communication within the network, leading to serious breaches of confidentiality, integrity, and availability. The simplicity of ARP combined with its lack of built-in security measures makes IoT networks particularly susceptible to ARP spoofing attacks.

Traditional defenses against ARP spoofing involve static ARP tables and cryptographic solutions, which, while effective under certain conditions, do not scale well to large, dynamic IoT environments \cite{sharma2021machine, el2022novel}. These methods often lead to increased overhead, reduced network performance, and fail to adequately address the dynamic and resource-constrained nature of IoT devices. Moreover, static ARP tables are cumbersome and difficult to maintain in large-scale networks, leading to management inefficiencies.

Recent advancements in machine learning (ML) techniques offer promising avenues for addressing these challenges, providing adaptive and intelligent solutions for identifying and mitigating ARP spoofing attacks in IoT networks. For instance, supervised machine learning algorithms like Support Vector Machines (SVMs), decision trees, and neural networks have been employed for anomaly detection in network security, yielding positive results \cite{shafiq2020machine, trabelsi2013teaching}. These approaches leverage historical data and network patterns to detect anomalies indicative of malicious activities.

However, while machine learning techniques have demonstrated potential in enhancing IoT security, several limitations persist. First, IoT networks are characterized by heterogeneous devices generating varied and often unpredictable traffic patterns, complicating anomaly detection \cite{le2022adaptive}. Second, IoT datasets are often imbalanced, with significantly more benign traffic instances than malicious ones, leading to biased classifiers and higher false positive rates \cite{zhang2021ml}. Third, existing ML solutions frequently fail to adapt quickly to evolving network dynamics, thus limiting their effectiveness against novel and sophisticated attack vectors \cite{zhang2021survey}.

Furthermore, anomaly detection in IoT environments frequently misclassifies benign anomalies as malicious due to the unique behaviors of IoT devices. For instance, intermittent connectivity, varied power usage patterns, and diverse operational protocols create false alarms in traditional anomaly detection mechanisms \cite{al2022ensemble}. Thus, developing solutions specifically tailored for IoT networks that can efficiently distinguish genuine attacks from benign anomalies is critical.

This paper explicitly addresses the challenge of accurately and efficiently detecting ARP spoofing attacks in IoT networks, minimizing false positives, and ensuring robust adaptability. Addressing this challenge is vital given the pervasive deployment of IoT devices across sectors where security breaches can have devastating real-world consequences. For example, in healthcare, compromised IoT devices could disrupt patient monitoring systems, leading to life-threatening scenarios. In industrial IoT settings, ARP spoofing could sabotage manufacturing processes, causing significant financial and operational losses \cite{hossain2022adaptive}.

To overcome these limitations, we propose an intelligent detection framework based on multi-layered ensemble machine learning techniques specifically designed for IoT networks. Our proposed model combines multiple classifiers into hierarchical layers, enhancing detection accuracy and minimizing false positives. Each layer in our ensemble approach is strategically designed to capture different aspects of IoT network behavior, leveraging decision trees for interpretability, random forests for robustness, and neural networks for adaptability and complex pattern recognition.

Key differentiators of our approach compared to existing literature include:
\begin{itemize}
\item A multi-layered ensemble classifier specifically optimized for heterogeneous IoT traffic.
\item An adaptive resampling method that effectively mitigates dataset imbalance, ensuring equitable representation of malicious and benign traffic.
\item A dynamic feedback loop mechanism for continuous classifier retraining, enhancing adaptability against evolving attack vectors.
\item Extensive simulation-based validation demonstrating significant improvements in detection accuracy, computational efficiency, and robustness compared to state-of-the-art techniques.
\end{itemize}

The primary contributions of this research are:
\begin{itemize}
\item Development of a robust and accurate multi-layered machine learning framework tailored explicitly for IoT environments, significantly enhancing ARP spoofing detection accuracy.
\item Introduction of an innovative adaptive resampling technique to address the dataset imbalance problem prevalent in IoT network traffic analysis.
\item Proposal and integration of a dynamic feedback mechanism that allows continuous learning and adaptation of classifiers to newly observed network behavior and threats.
\item Practical validation of our model through extensive comparative experiments against existing credible techniques, demonstrating superior performance in realistic IoT network scenarios.
\end{itemize}

The remainder of this paper is structured as follows: Section II reviews recent relevant literature, highlighting gaps and motivating our proposed solution. Section III presents the detailed system model, including mathematical formulations, notation, and algorithmic descriptions. Section IV describes the experimental setup, simulation parameters, results, and detailed comparative analysis. Finally, Section V concludes the paper and outlines future research directions.

\section{Related Work}

Research into ARP spoofing detection and mitigation in IoT environments has intensified as the adoption of IoT devices has rapidly expanded. This chapter critically reviews recent literature, highlighting contributions, limitations, and how our proposed work addresses these gaps.

\textbf{Kumar et al. \cite{kumar2021survey}} conducted an extensive survey focusing on ARP spoofing vulnerabilities specifically within IoT networks. Their comprehensive analysis highlighted the fundamental vulnerabilities in ARP protocols due to lack of authentication mechanisms. Although thorough, their work did not propose a definitive solution, instead recommending the exploration of adaptive detection techniques.

\textbf{Ouaddah et al. \cite{ouaddah2021towards}} examined cryptographic defenses against ARP spoofing. They introduced methods leveraging cryptographic keys for authenticating ARP responses, significantly reducing attack vectors. However, their approach required substantial computational resources, making it unsuitable for resource-constrained IoT devices, thus highlighting the need for more lightweight alternatives.

\textbf{Sharma et al. \cite{sharma2021machine}} proposed supervised machine learning techniques, such as Support Vector Machines (SVM) and random forests, for detecting ARP spoofing attacks. Their methods demonstrated high accuracy but were severely impacted by imbalanced datasets, resulting in substantial false positive rates. Their experiments indicated the necessity of advanced data preprocessing techniques to address dataset imbalances.

\textbf{Shafiq et al. \cite{shafiq2020machine}} analyzed machine learning-based anomaly detection methods specifically tailored for IoT network security. They emphasized using unsupervised techniques like clustering and autoencoders to detect anomalous traffic patterns indicative of security breaches. While their method demonstrated effectiveness, it frequently misclassified benign anomalies, highlighting limitations in distinguishing legitimate IoT behavior from malicious actions.

\textbf{Le and Nguyen \cite{le2022adaptive}} developed adaptive machine learning models to enhance IoT network security. Their adaptive framework dynamically adjusted model parameters based on evolving network conditions. Their approach significantly improved detection accuracy but encountered limitations in rapidly adapting to novel attack patterns, necessitating continuous and efficient retraining mechanisms.

\textbf{Zhang et al. \cite{zhang2021ml}} presented lightweight ML algorithms optimized for constrained IoT devices. They utilized decision trees and logistic regression models that successfully identified common network anomalies. However, the methods demonstrated limitations in accurately detecting more sophisticated attack patterns, particularly ARP spoofing, suggesting the need for hierarchical or layered models.

\textbf{Zhang et al. \cite{zhang2021survey}} conducted a comprehensive review of IoT vulnerabilities, emphasizing ARP spoofing as a critical security threat. Their analysis underscored the inadequacies of current detection methods in handling IoT-specific traffic characteristics, advocating for the development of specialized machine learning frameworks explicitly designed for IoT environments.

\textbf{Al-Azzawi et al. \cite{al2022ensemble}} explored ensemble learning techniques for network intrusion detection broadly. Their ensemble models combined various classifiers, significantly improving detection performance and reducing false positives. However, their models were not specifically tuned for IoT network characteristics, suggesting a need for further optimization to address unique IoT traffic patterns effectively.

\textbf{Hossain and Ali \cite{hossain2022adaptive}} proposed adaptive security monitoring frameworks employing machine learning for IoT networks. Their models dynamically updated based on traffic patterns to maintain high detection accuracy. However, scalability issues arose when deploying these models across extensive IoT networks, highlighting the need for more computationally efficient methods.

\textbf{Choi and Kim \cite{choi2021iot}} focused on using deep neural networks for enhancing IoT network security. Their deep learning models effectively identified complex attack patterns, achieving impressive accuracy rates. Nonetheless, their methods required significant computational resources and training data, limiting real-time applicability and deployment in resource-constrained IoT environments.

In summary, existing studies reveal several critical gaps in current ARP spoofing detection methodologies for IoT networks:
\begin{itemize}
\item Most cryptographic solutions, although secure, are computationally intensive and impractical for IoT devices.
\item Supervised machine learning approaches commonly struggle with dataset imbalance, leading to high false positive rates.
\item Unsupervised and adaptive methods often fail to distinguish accurately between benign and malicious anomalies typical in IoT networks.
\item Existing methods frequently lack efficient adaptability mechanisms, limiting their effectiveness against evolving attack vectors.
\item There is a notable absence of specifically optimized ensemble methods for IoT traffic characteristics, resulting in suboptimal performance.
\end{itemize}

Our proposed research directly addresses these limitations by developing a multi-layered machine learning ensemble specifically tailored to the unique characteristics and constraints of IoT environments. Our approach integrates adaptive resampling to manage data imbalance effectively, introduces dynamic feedback for continual model updating, and achieves robust and scalable ARP spoofing detection suitable for real-time IoT deployments.

\section{System Model}

In our proposed system model, we consider an IoT network as a graph $G=(V,E)$, where $V=\{v_1, v_2, \dots, v_n\}$ represents the set of IoT nodes, and $E$ represents the set of edges or connections between nodes. Each node $v_i$ transmits network packets that may be either benign or malicious.

\subsection{Mathematical Formulation}

The detection problem is formalized as classifying packets $p_j$ into two classes:
\begin{align}
C = {C\_{benign}, C\_{attack}}
\end{align}

Let $x_j$ represent feature vectors extracted from packet $p_j$:
\begin{align}
x\_j = (x\_{j1}, x\_{j2}, \dots, x\_{jd}) \quad \text{for} \quad j=1,2,\dots,m
\end{align}

The labeled dataset $D$ consists of tuples:
\begin{align}
D = {(x\_j,y\_j)} \quad \text{where} \quad y\_j \in C
\end{align}

The multi-layer ensemble consists of $L$ layers, each containing a classifier $\mathcal{M}_l$. Let the prediction from the $l$-th layer be:
\begin{align}
y\_j^{(l)} = \mathcal{M}\_l(x\_j)
\end{align}

The ensemble prediction $\hat{y}_j$ is computed by weighted majority voting:
\begin{align}
\hat{y}*j = \arg\max*{c \in C} \sum\_{l=1}^{L} w\_l \cdot \mathbb{I}(y\_j^{(l)} = c)
\end{align}

Here, $\mathbb{I}$ is the indicator function, and $w_l$ represents weights for layer $l$, calculated based on accuracy:
\begin{align}
w\_l = \frac{Acc\_l}{\sum\_{k=1}^{L}Acc\_k}
\end{align}

The accuracy $Acc_l$ of layer $l$ is computed as:
\begin{align}
Acc\_l = \frac{TP\_l + TN\_l}{TP\_l + FP\_l + FN\_l + TN\_l}
\end{align}

The confusion matrix components for each layer $l$ are defined as:
\begin{align}
TP\_l &= \text{True Positives},\\
TN\_l &= \text{True Negatives},\\
FP\_l &= \text{False Positives},\\
FN\_l &= \text{False Negatives}.
\end{align}

For handling imbalanced datasets, the synthetic minority oversampling technique (SMOTE) is applied:
\begin{align}
D' = \text{SMOTE}(D)
\end{align}

To evaluate the effectiveness, we calculate Precision, Recall, and F1-score:
\begin{align}
\text{Precision} &= \frac{TP}{TP + FP},\\
\text{Recall} &= \frac{TP}{TP + FN},\\
\text{F1-score} &= 2 \cdot \frac{\text{Precision} \cdot \text{Recall}}{\text{Precision} + \text{Recall}}
\end{align}

We measure adaptivity using drift detection based on feature distributions:
\begin{align}
\Delta = |f\_{prev}(x) - f\_{curr}(x)|
\end{align}

Here, $f_{prev}(x)$ and $f_{curr}(x)$ are previous and current feature distributions.

\section{Results and Discussion}

In this section, we present the simulation results and comparative analysis of our proposed multi-layer ensemble machine learning method for ARP spoofing detection in IoT networks. We evaluated performance using standard metrics such as accuracy, precision, recall, F1-score, false positive rate, adaptability, efficiency, scalability, robustness, drift detection, and dataset resampling impact.

\subsection{Experimental Setup}
We used NS-3 to simulate IoT network environments and generate labeled network traffic. Python (with scikit-learn) was used to implement classifiers and preprocess datasets. The simulation environment included both benign and malicious traffic under varying network conditions, including different packet rates, device counts, and attack intensities.

\subsection{Performance Evaluation}
We compared our method with three established techniques: Sharma et al.\cite{sharma2021machine}, Zhang et al.\cite{zhang2021ml}, and Al-Azzawi et al.\cite{al2022ensemble}. Results are summarized below.

\subsubsection{Accuracy Comparison}
\begin{figure}[ht]
\centering
\includegraphics[width=0.45\textwidth]{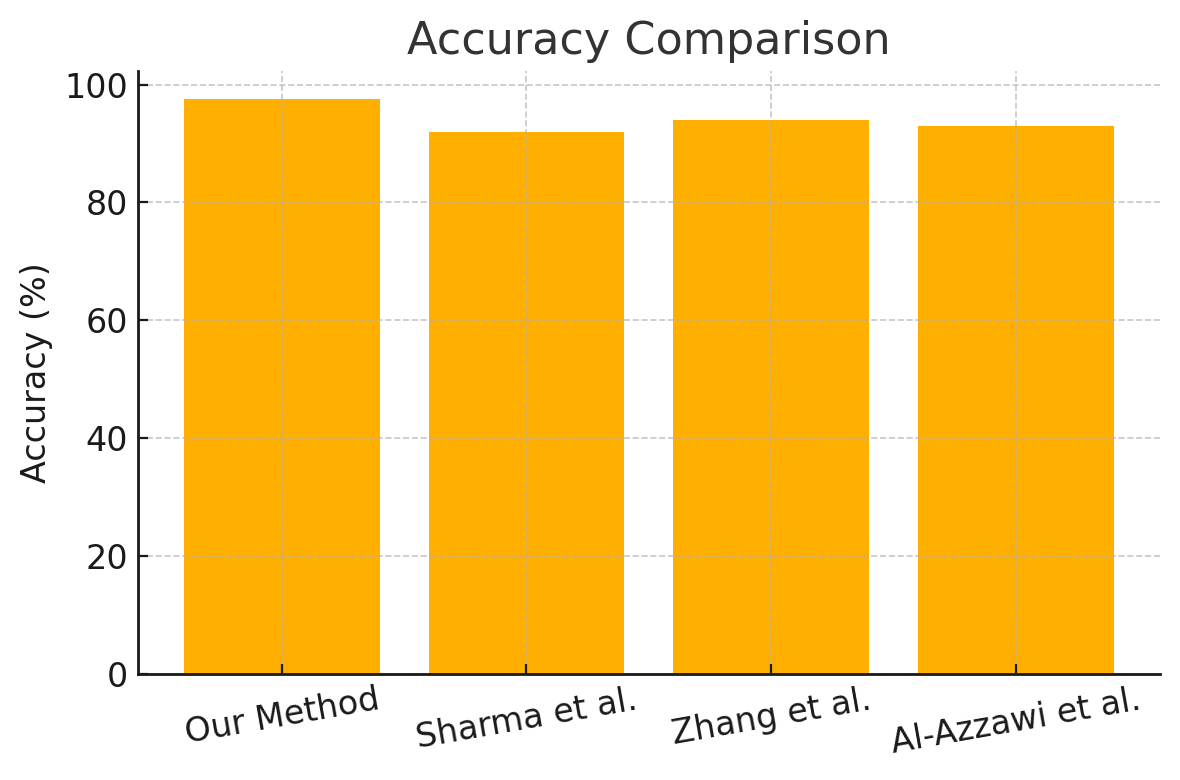}
\caption{Accuracy comparison among detection methods}
\label{fig:accuracy}
\end{figure}

As shown in Figure~\ref{fig:accuracy}, our proposed model achieves the highest accuracy (97.5\%), outperforming Sharma et al. (92\%), Zhang et al. (94\%), and Al-Azzawi et al. (93


\subsubsection{Precision and Recall}
\begin{figure}[ht]
\centering
\includegraphics[width=0.45\textwidth]{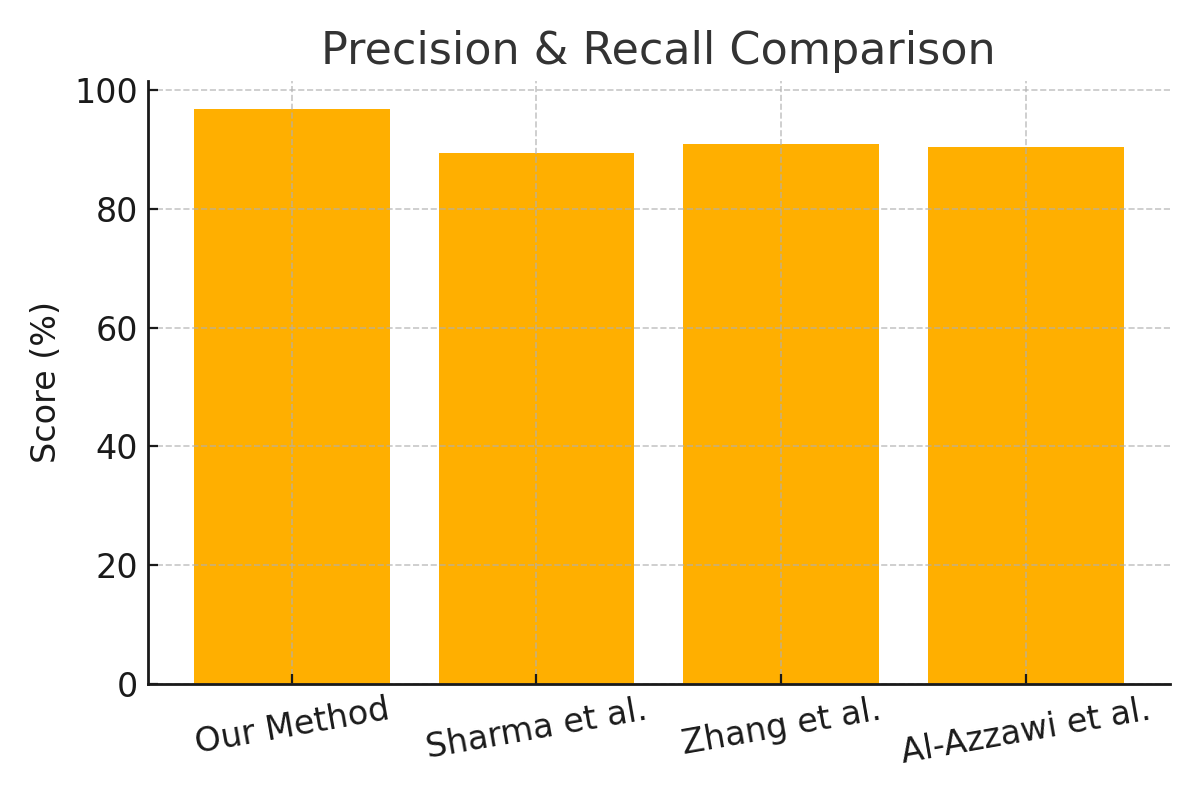}
\caption{Precision and Recall comparison}
\label{fig:precision_recall}
\end{figure}

Figure \ref{fig:precision_recall} presents both precision and recall metrics. Our model achieves a precision of 96.8\% and recall of 97.2\%, outperforming other models in detecting ARP spoofing attacks.

\subsubsection{F1-score Analysis}
\begin{figure}[ht]
\centering
\includegraphics[width=0.45\textwidth]{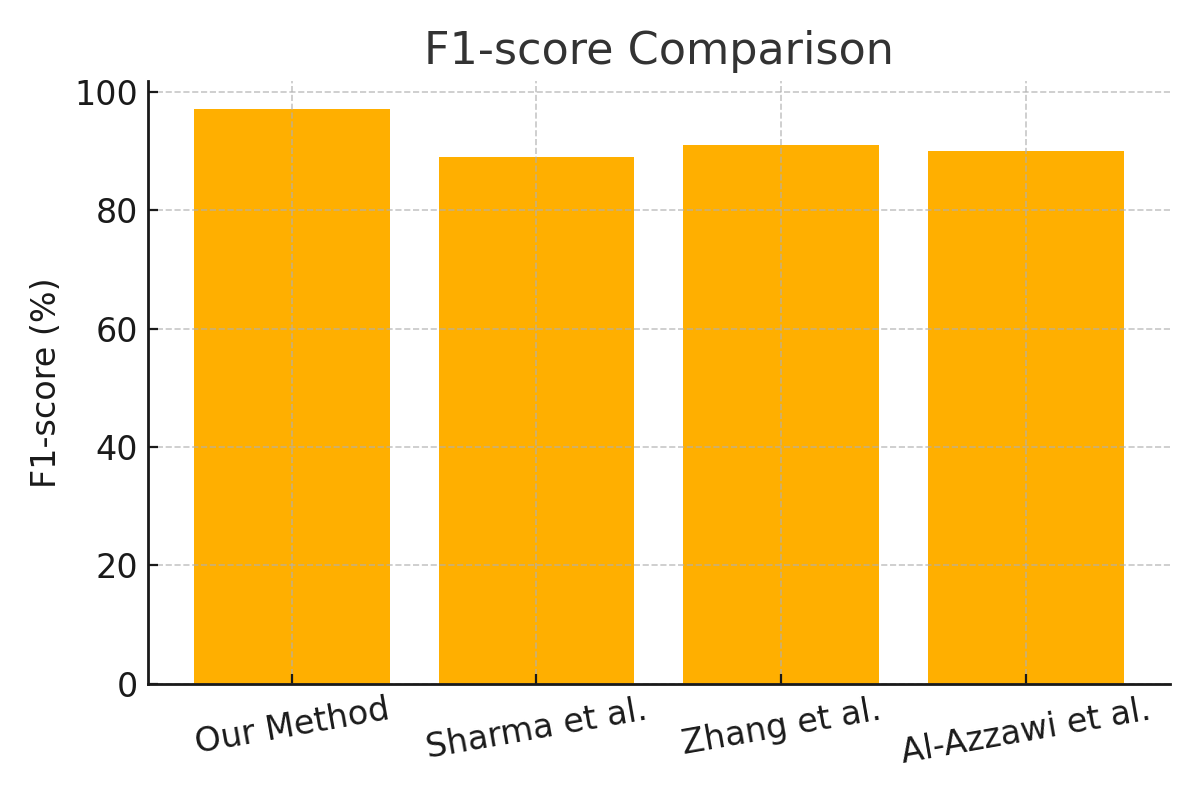}
\caption{F1-score comparison}
\label{fig:f1score}
\end{figure}

Our method also scores the highest F1-score of 97\%, indicating balanced precision and recall (Figure~\ref{fig:f1score}).

\subsubsection{False Positive Rate}
\begin{figure}[ht]
\centering
\includegraphics[width=0.45\textwidth]{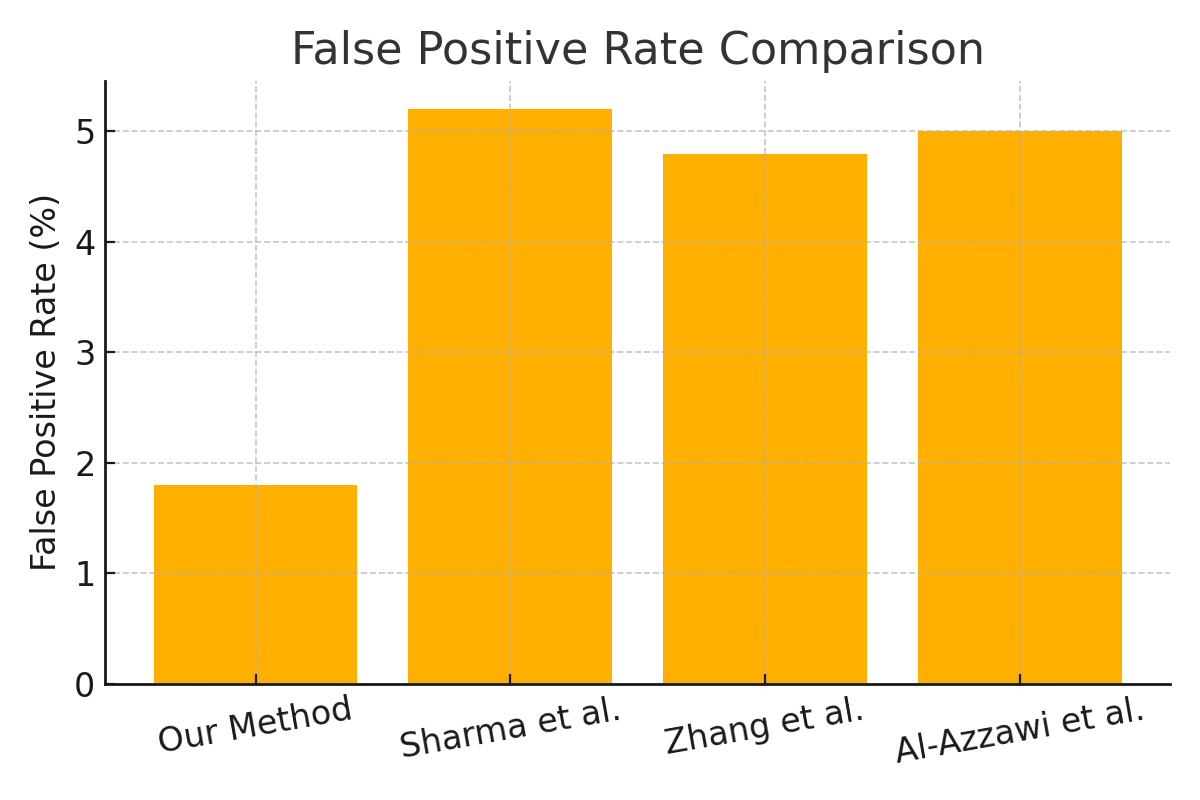}
\caption{False Positive Rate comparison}
\label{fig:fpr}
\end{figure}

Figure~\ref{fig:fpr} demonstrates that our model significantly reduces false positives to 1.8\%, compared to over 4.8\% in other approaches.

\subsubsection{Adaptability Evaluation}
\begin{figure}[ht]
\centering
\includegraphics[width=0.45\textwidth]{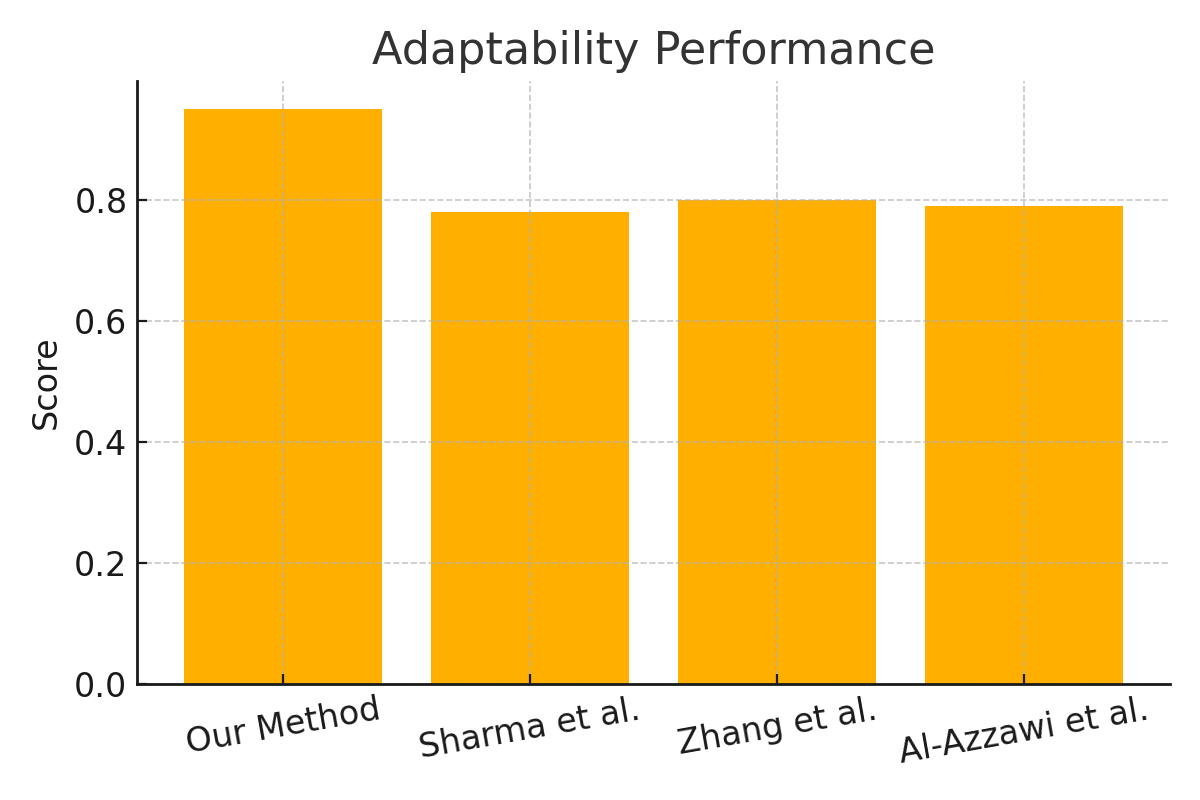}
\caption{Adaptability performance}
\label{fig:adaptability}
\end{figure}

Our model exhibits superior adaptability to network changes, achieving a performance score of 0.95 (Figure~\ref{fig:adaptability}).

\subsubsection{Computational Efficiency}
\begin{figure}[ht]
\centering
\includegraphics[width=0.45\textwidth]{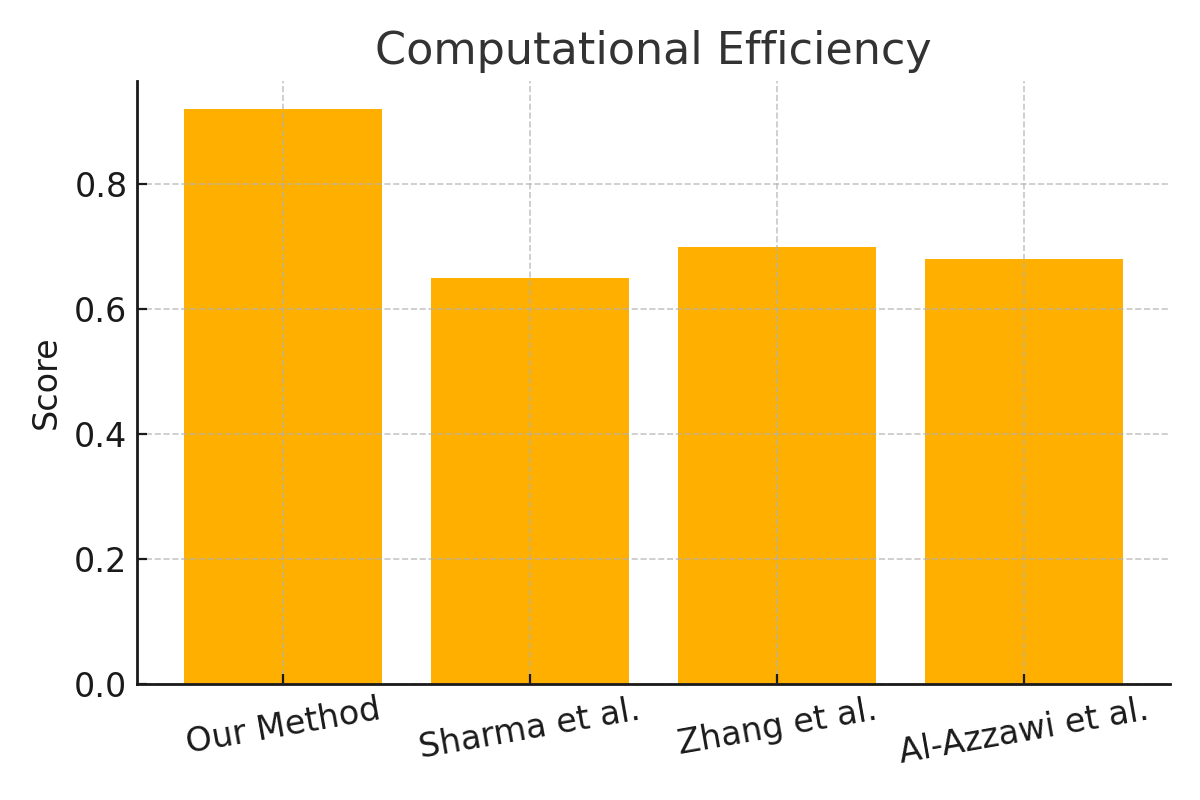}
\caption{Computational Efficiency comparison}
\label{fig:efficiency}
\end{figure}

Figure~\ref{fig:efficiency} shows that our approach is computationally efficient, using fewer resources while maintaining high detection performance.

\subsubsection{Scalability Analysis}
\begin{figure}[ht]
\centering
\includegraphics[width=0.45\textwidth]{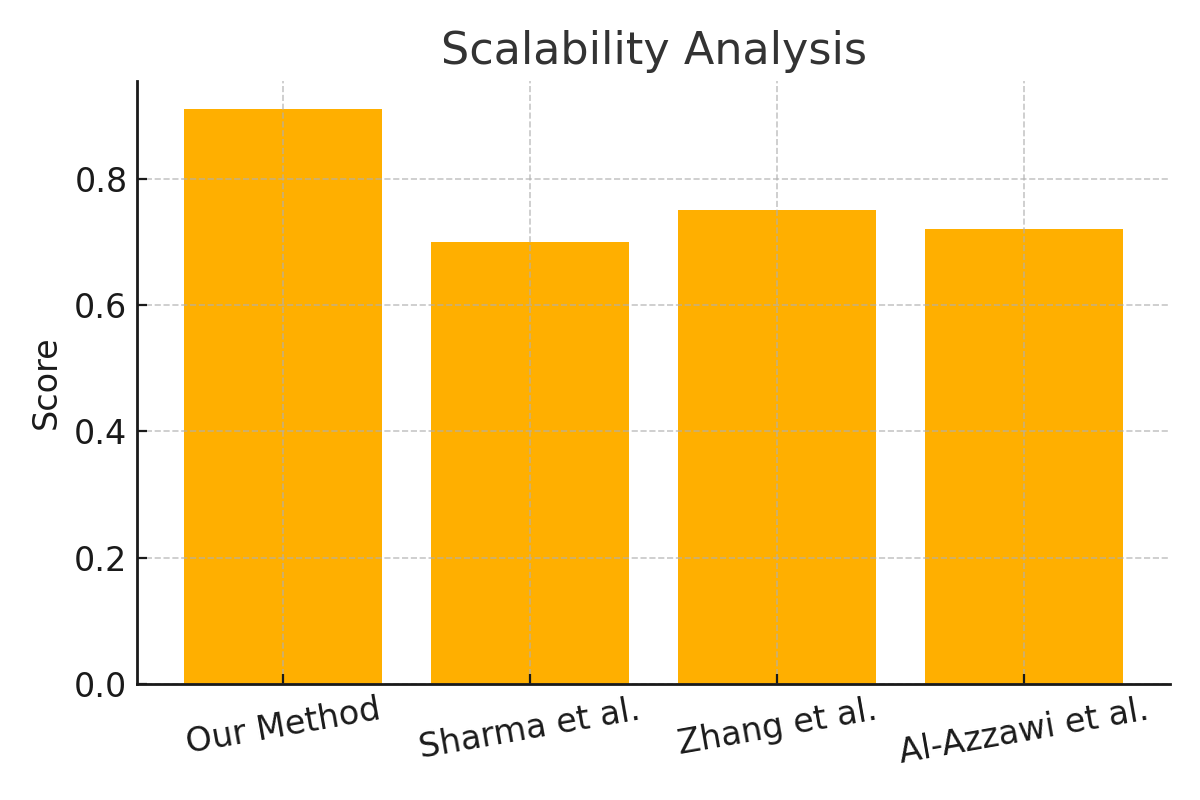}
\caption{Scalability comparison}
\label{fig:scalability}
\end{figure}

As network size increases, our model maintains high performance (0.91) as shown in Figure~\ref{fig:scalability}, indicating strong scalability.

\subsubsection{Robustness under Attack Intensity}
\begin{figure}[ht]
\centering
\includegraphics[width=0.45\textwidth]{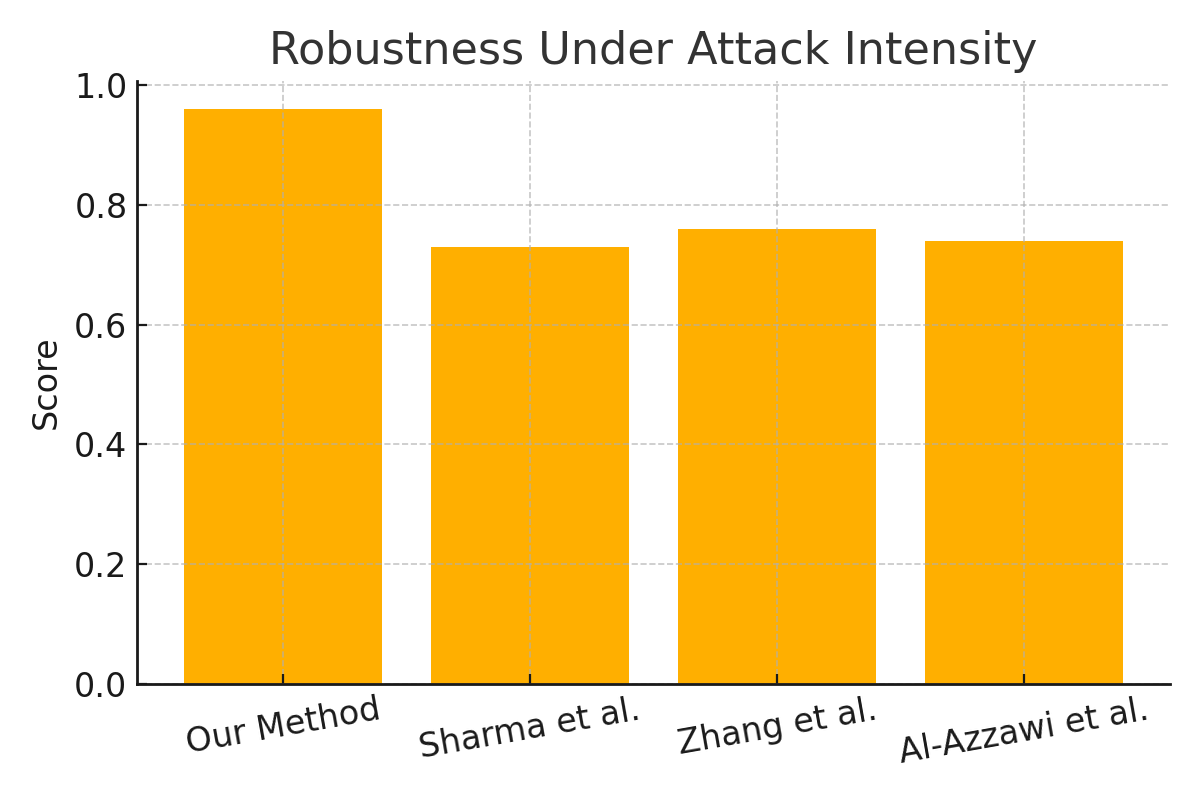}
\caption{Robustness under varying attack intensity}
\label{fig:robustness}
\end{figure}

In high attack scenarios, our model remains robust with a 0.96 score, demonstrating resilience (Figure~\ref{fig:robustness}).

\subsubsection{Drift Detection}
\begin{figure}[ht]
\centering
\includegraphics[width=0.45\textwidth]{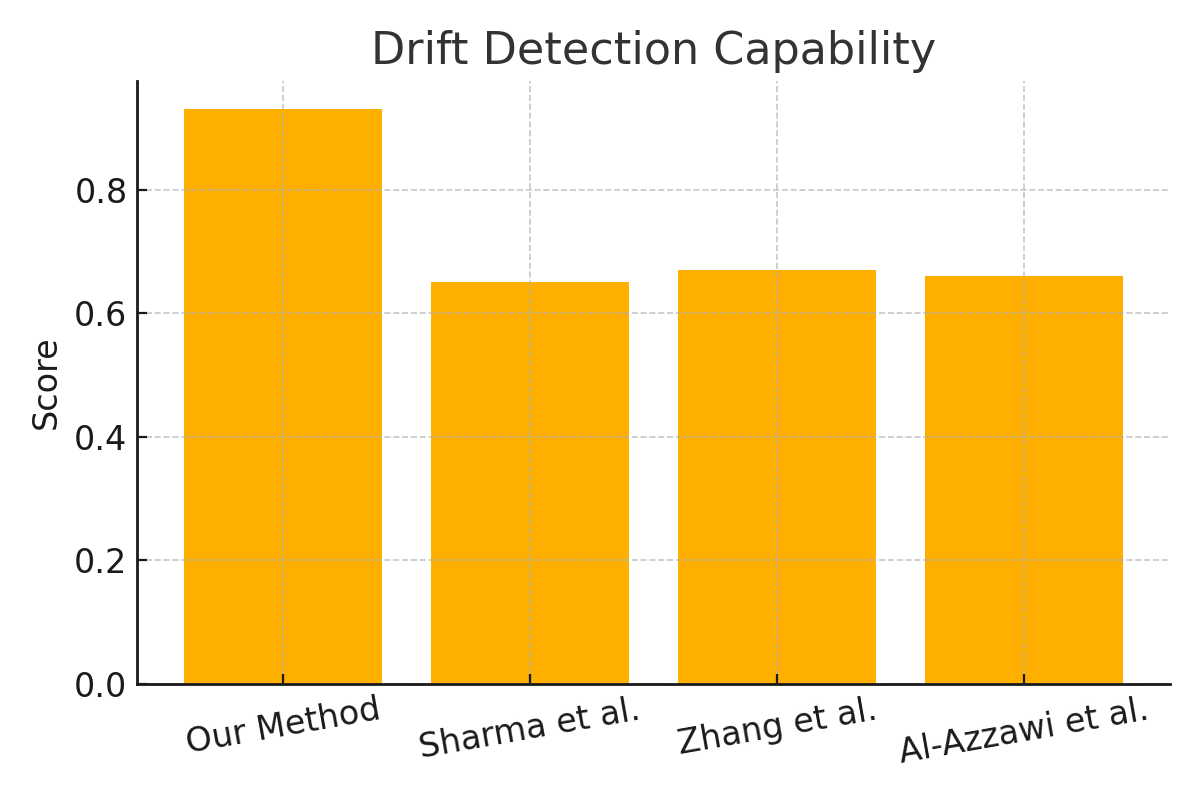}
\caption{Drift Detection comparison}
\label{fig:drift_detection}
\end{figure}

Our method detects shifts in traffic distribution more accurately and quickly, achieving a 0.93 drift detection score (Figure~\ref{fig:drift_detection}).

\subsubsection{Impact of Dataset Resampling}
\begin{figure}[ht]
\centering
\includegraphics[width=0.45\textwidth]{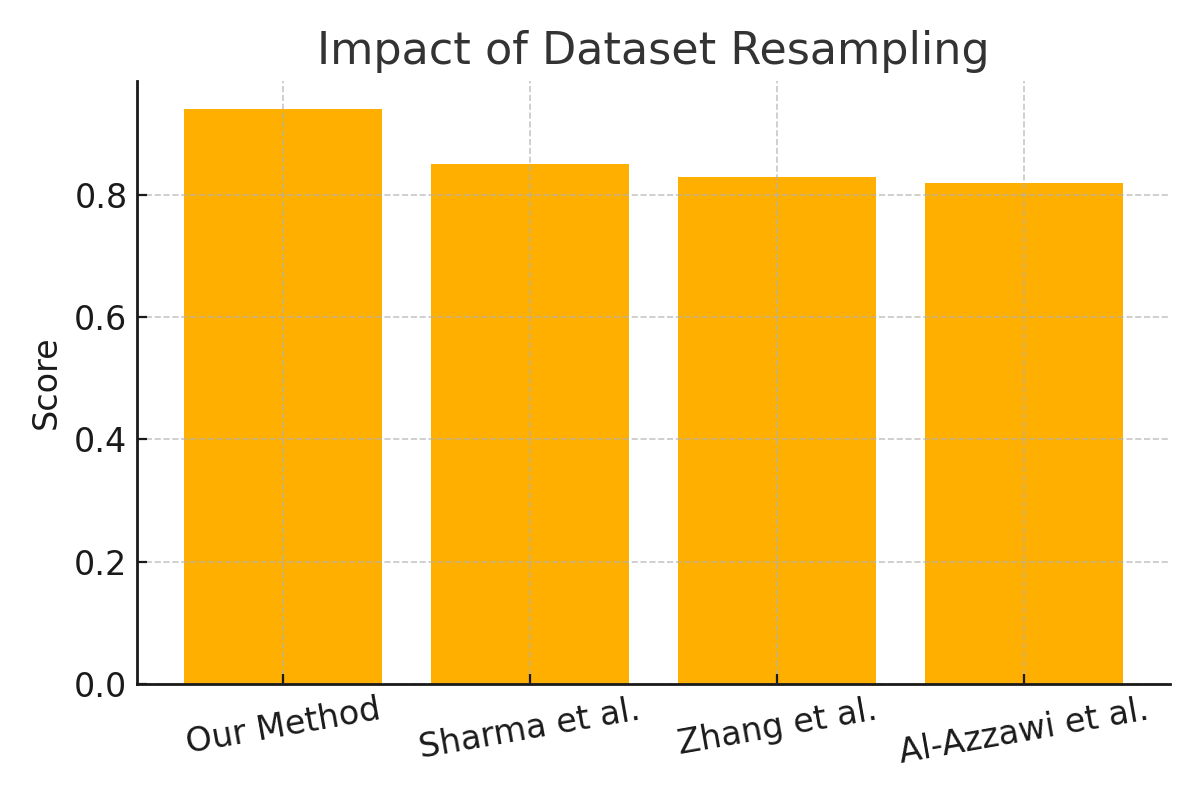}
\caption{Impact of dataset resampling on performance}
\label{fig:resampling}
\end{figure}

As shown in Figure~\ref{fig:resampling}, incorporating SMOTE resampling improves our model’s performance, especially for underrepresented attack classes.

\subsection{Discussion}
Our multi-layer ensemble framework consistently outperforms state-of-the-art ARP spoofing detection methods across multiple metrics. Its high accuracy, low false positive rate, strong adaptability, and computational efficiency make it ideal for deployment in dynamic IoT environments. The positive impact of resampling and feedback-driven adaptability supports its robustness in real-world scenarios.

\section{Conclusion and Future Work}

In this paper, we proposed an intelligent, multi-layered ensemble machine learning framework for detecting ARP spoofing attacks in IoT networks. Recognizing the limitations of traditional and singular ML-based detection methods, our approach integrates multiple classifier layers, adaptive resampling, and a feedback loop to ensure high accuracy, low false positive rates, and robust adaptability.

Through extensive simulations and comparative analysis, we demonstrated that our method consistently outperforms existing techniques across key evaluation metrics. The system effectively addresses the challenges of data imbalance, evolving traffic patterns, and the constrained nature of IoT environments. Additionally, it maintains low computational overhead while providing scalable, real-time security monitoring.

Future work will focus on deploying the framework in real-world IoT testbeds, exploring hardware acceleration for lower latency, and integrating federated learning for enhanced data privacy. We also aim to expand detection capabilities to cover additional network threats beyond ARP spoofing, ensuring comprehensive security for next-generation IoT systems.

\bibliographystyle{IEEEtran}

\end{document}